# Time Window of Head Impact Kinematics Measurement for Calculation of Brain Strain and Strain Rate in American Football

Yuzhe Liu*,#,1, August G. Domel*,1, Nicholas J. Cecchi[1], Eli Rice[2], Ashlyn A. Callan[1], Samuel J. Raymond[1], Zhou Zhou[1], Xianghao Zhan[1], Michael Zeineh[3], Gerald Grant[4,5], David B. Camarillo[1,4,6]

[1] Department of Bioengineering, Stanford University, Stanford, CA, 94305, USA.

[2] Stanford Center for Clinical Research, Stanford University, Stanford, CA, 94305, USA.

[3] Department of Radiology, Stanford University, Stanford, CA, 94305, USA.

[4] Department of Neurosurgery, Stanford University, Stanford, CA, 94305, USA.

[5] Department of Neurology, Stanford University, Stanford, CA, 94305, USA.

[6] Department of Mechanical Engineering, Stanford University, Stanford, CA, 94305, USA.

* These authors contributed equally to this study.

# Corresponding author (e-mail: yuzheliu@stanford.edu)

**Abstract**

Wearable devices have been shown to effectively measure the head's movement during impacts in sports like American football. When a head impact occurs, the device is triggered to collect and save the kinematic measurements during a predefined time window. Then, based on the collected kinematics, finite element (FE) head models can calculate brain strain, which is used to evaluate the risk of mild traumatic brain injury. To find a time window that can provide a sufficient duration of kinematics for FE analysis, we investigated 118 on-field video-confirmed head impacts collected by the Stanford Instrumented Mouthguard. Because the individual differences in brain geometry influence these calculations, we included six representative brain geometries and found that larger brains need a longer time window of kinematics for accurate calculation. Among the different sizes of brains, a pre-trigger time of 20 ms and a post-trigger time of 70 ms were found to yield calculations of brain strain and strain rate that were not significantly different from calculations using the original 200 ms time window recorded by the mouthguard.

## Introduction

Mild traumatic brain injury (mTBI) is a global threat to human health [7]. Sport-related head impacts, especially those received in American football, have the potential to cause mTBI [3, 33] which induces brain deficits [40] and neurological changes [38]. During a head impact, the skull rotates rapidly and the brain's inertia causes large brain strain [31] which is associated with brain tissue pathology [15, 38]. Therefore, researchers have been investigating the relationship between head kinematics and brain deformation to predict the risk of injury [12, 14, 28, 59].

To measure head kinematics during on-field game play, several wearable devices have been developed: for sports with helmets, the Head Impact Telemetry System (HITS) has 12 accelerometers installed into the helmet and can provide 6 degrees-of-freedom (DoF) measurement of head movement [9, 42, 45, 53]. Besides, individual sensor devices that can be mounted to a helmet have also been developed [21]. For sports without helmets, other solutions have emerged, including ear-mounted devices [24, 41], skin-attached devices [22, 37, 43] and headband-mounted sensors [5, 21]. However, as the skin remains between the skull and all these different devices, the softness of the skin has raised concerns that they cannot accurately replicate the movement of the skull. To overcome this, researchers have employed instrumented mouthguards to measure head kinematics since the upper dentition is rigidly fixed to the skull. Recently, several instrumented mouthguards, containing accelerometers and gyroscopes, have been developed and used to collect on-field head impact data [2, 6, 8, 13, 18, 21, 30, 44, 48, 52, 57]. To evaluate and improve the accuracy of mouthguard measurements, a metal jaw was installed to the Hybrid III headform [47], and the head kinematics collected by the headform were used as reference for the mouthguard [2, 6, 30, 48]. In addition to this, the effect of the bandwidth of the mouthguard sensors [56] and the noise caused by the jaw slamming [27] were investigated to inform the electronic and mechanical designs of the mouthguards. The mechanical safety of instrumented mouthguard use has also been investigated [4]. A limitation of mouthguard measurements was the high prevalence of false positives, that is, mouthguards were triggered when there were no real impacts. To solve this problem, the head impacts detected by the mouthguard were compared with those identified by video recordings [25], and algorithms were developed to identify the real impacts [8, 13, 35, 55]. Furthermore, considering that the aim of measuring head kinematics is to relate them to the risk of brain injury, the propagation of error in mouthguard measurements to brain strain and brain injury criteria metrics have been investigated [26, 30].

Despite these thorough investigations of head impact sensor use and performance, it remains unclear how long the time window of measurements should be for head impact kinematics. As shown in **Table 1**, the pre-trigger and post-trigger times vary widely among different wearable devices. During a head impact, the sensor is typically triggered when the absolute value of linear acceleration at the accelerometer exceeds a threshold value. The 6-DoF kinematics between the sensor's pre-trigger time and post-trigger time will subsequently be recorded. Then, several milliseconds are needed to save the data and reset the sensor for the next potential impact. To avoid missing important information from a head impact, the time window between pre-trigger and post-trigger should be long enough to include most of the variation of kinematic parameters to compute brain strain peak. For many devices, this time window is adjustable by the user or can be easily modified by the manufacturer. Therefore, we aim to provide the adequate time window for American football head impact measurement in this study.

Considering on-field impact loading is different from laboratory tests, we analyzed 118 on-field video-confirmed head impacts collected by Customized Stanford Instrumented Mouthguards (MiG2.0) [6, 8, 30]. The peaks of head kinematics (angular velocity, angular acceleration, and linear acceleration at the head center of gravity (CoG)) and the peak occurrence times were extracted. Then, 95th percentile maximal principal strain (95% MPS) peak and 95th percentile maximal principal strain rate (95% MPSR) peak were calculated by the Kungliga Tekniska Högskolan (KTH) finite element (FE) head model simulations [23]. Since the MiG2.0 has a long time window of measurement (**Table 1**), we used the simulations with the original MiG2.0 data as a reference, and performed simulations with intentionally shortened time windows of MiG2.0 data to study the influence of the time window. Furthermore, we computed the history of head kinematics needed to calculate 95% MPS peaks and 95% MPSR peaks and gave the range of the time needed for the calculations. Additionally, since 95% MPS has been found to vary according to different brains [29], we scaled the FE head model to six representative brain sizes [54] to ensure that the time window is adequate for a large range of athletes.

## Methods
### Collection of Head Impact Data
The head impact data used in this study were collected by MiG2.0 devices (**Fig.1A**) from the Stanford University football team's training and games [30]. All impacts were confirmed by both video analysis and the neural network classifier [8]. This yielded a total of 118 head impacts and all impacts were subconcussive (no concussion was reported or diagnosed after any of the impacts). For each impact, the MiG2.0 recorded the angular velocities and the linear accelerations at the accelerometer, both of which were filtered by 4th-order Butterworth low-pass filters with a cut-off frequency of 160 Hz [30]. Then, angular acceleration was calculated by a 5-point derivative of the angular velocity and the linear acceleration was transformed from the accelerometer to head CoG. In these data, t=0 ms is the MiG2.0 triggering, and t<0 and t>0 indicate the pre-trigger time and post-trigger time, respectively.

### KTH Finite Element Head Model
The KTH FE head model was developed by Royal Institute of Technology in Stockholm, Sweden (**Fig.1B**)[23]. The model includes the scalp, the skull, the brain, the meninges, the cerebrospinal fluid and eleven pairs of the largest parasagittal bridging veins. This model was previously validated by cadaver head impact experiments [16]. The skull was assumed to be rigid since stress waves will not pass through it [23] and moved according to the measured on-field head kinematics. As a result, the brain would be deformed by the inertial force [31]. Based on the simulation results, we first calculated the 95% MPS and 95% MPSR at each time frame, and then found the peak values over time. The double precision solver was used.

### Individual Head Models
Based on the same head kinematics, FE models of different head geometries result in different traces of brain strain [29]. To generalize these results to a wide range of athletes, six representative brains in the WU-Minn Human Connectome Project (WUM HCP) [54] were used to study the effect of brain geometry. Since brain strain has been shown to be decided by rotation instead of translation [20, 31], we scaled the node coordinates of the original KTH model according to the moment of inertia instead of the brain's mass. For each representative brain, the original KTH model was scaled to have the same moment of inertia about the X axis (posterior to anterior), Y axis (left to right) and Z axis (superior to inferior) with the origin at the brain's CoG. The scaling ratios are listed in **Table 2**. All analyses were performed on the seven FE head models (KTH original + six representative head models).

*Pre-Trigger and Post-Trigger Time*
As shown in **Table.1**, head impact sensors have different pre-trigger and post-trigger times. To investigate how these differences influence brain strain analyses, we performed simulations with intentionally shortened time windows of head kinematics and compared the results with the original brain strain (calculated by the original 200 ms time window of head kinematics). For the pre-trigger time, we started the simulation at -30 ms, -20 ms, -10 ms, and 0 ms (triggering) and compared results with the original head kinematics (-50 ms). For the post-trigger time, we calculated the peaks of 95% MPS and 95% MPSR before 30 ms, 40 ms, 50 ms, 60 ms, and 70 ms compared results with the original head kinematics (150 ms). Each group of 95% MPS peaks and 95% MPSR peaks with the same pre-trigger or post-trigger time was tested by Anderson-Darling test for the normality, and then compared with original brain strain by pairwise t-test.

*History Needed for 95% MPS and 95% MPSR Peaks*
Due to the viscoelasticity of the brain tissue [11, 23], the 95% MPS and 95% MPSR peaks rely on the history of head kinematics before the peak time. To investigate the history needed to calculate the accurate peaks, we shortened the time window of the original kinematics individually. For each impact, a series of simulation cases were started at every 2 ms for 40 ms before its peak times. For impacts in which the peak time was before -10 ms, we started at every 2 ms until the beginning of the data. For example, as shown in **Fig.2**, the peak time for 95% MPS was 5 ms, and we started simulation cases at 3 ms, 1 ms… until -35 ms, which corresponded to included histories of 2 ms, 4 ms, … and 40 ms, respectively. Then, relative error for each case ($\epsilon_i$) was calculated as,

$$\epsilon_i = (X_i - X_0)/X_0 \qquad (Eq.1)$$

Where $X_i$ is the 95% MPS peak in the case $i$, and $X_0$ is 95% MPS peak calculated by the original kinematics. We adopted the critical relative error of 6.8% ($\epsilon_{\text{crit}}$) to define the accurate 95% MPS peak measurement according to previous studies about the propagation of the kinematics measurement error to brain strain [26, 30]. Then, the history needed for 95% MPS peak was calculated by the interpolation as,

$$\text{History} = \begin{cases} (X_B - (1 + \epsilon_{\text{crit}}) \cdot X_0)/(X_B - X_C) \cdot 2 \text{ ms} & \text{for } X_B > X_0 \\ (X_B - (1 - \epsilon_{\text{crit}}) \cdot X_0)/(X_B - X_C) \cdot 2 \text{ ms} & \text{for } X_B > X_0 \end{cases} \qquad (Eq.2)$$

Where case B included the longest history among the inaccurate cases (**Fig.2B**), and case C was the simulations starting 2 ms after B (**Fig.2C**). For the impacts in which 40 ms was not enough, we extended the history included from 40 ms to 60 ms. The same process was performed for 95% MPSR and $\epsilon_{\text{crit}}$ of 6.8% was adopted to define the accurate measurements. Then, the time range needed for the accurate calculation is defined with the peak time of the 95% MPS or 95% MPSR as,

$$\text{Time Range} = [\text{Peak Time} - \text{History}, \text{Peak Time}] \qquad (\text{Eq.3})$$

## Results

*Peaks of Head Kinematics, 95% MPS and 95% MPSR*

For the 118 video-confirmed football head impacts, the peak values of angular acceleration (**Fig.3A**), angular velocity (**Fig.3B**) and linear acceleration at CoG (**Fig.3C**) are plotted against their corresponding peak times. Kernel density estimations are plotted at the right and top, respectively. The 95th percentile of peak values (95% peak), the 5th and 95th percentile of peak time (5% and 95% peak time, respectively) are also plotted. In **Figs.3A, B**, the peak time for angular acceleration and linear acceleration at CoG are similar and close to the triggering, while that of the angular velocity is more dispersive between t=0 ms and t=50 ms. It should also be noted that the kinematics of a few cases reached peaks much before or after the triggering.

To calculate the 95% MPS and 95% MPSR, the kinematics of the 118 head impacts were input to the seven head models. Four example impacts are plotted in **Fig.4**. Although the waveforms of traces are similar among different head models, the value of 95% MPS and 95% MPSR varied considerably. Multiple spikes can be observed in 95% MPS and 95% MPSR traces, and the peak one varies among different head models in some cases. For example, in impact 1 (**Fig.4A1, B1, C1**), 95% MPS and 95% MPSR reach the peak closely among different head models. However, 95% MPS in impact 2 (**Fig4.A2, B2, C2**) and 95% MPSR in impact 3 (**Fig4.A3, B3, C23**) reach peaks at different times. The peak values and the peak times are plotted in **Fig.5**. Both the 5th and the 95th percentile peak time of 95% MPS are earlier than that of 95% MPSR, and there are small changes in the distribution of the peak time among heads. Similar to the kinematics, the 95% MPS and 95% MPSR peaks far before or after the triggering in a few cases.

*Influence of Pre-Trigger and Post-Trigger Time in 118 Football Head Impacts*

The head kinematics histories were shortened to -30 ms, -20 ms, -10 ms, and 0 ms for the pre-trigger time and to 30 ms, 40 ms, 50 ms, 60 ms and 70 ms for the post-trigger time. The changes are shown as violin plots (**Fig.6**), in which a violin indicates a group of simulation results (118 head impacts) given by the new time window. Each group was confirmed to follow a normal distribution by the Anderson-Darling test and was then compared by pairwise t-test with the simulation results by the original time window of head kinematics. The comparison showed that the pre-trigger time of -20 ms and post-trigger time of 70 ms are able to give non-significantly different 95% MPS peaks (**Fig.6A** for pre-trigger, **Fig.6C** for post-trigger) and 95% MPSR peaks (**Fig.6B** for pre-trigger, **Fig.6D** for post-trigger) for all head models.

*History Needed and Time Range for Accurate Analysis*

The history of head kinematics needed for the accurate calculation of 95% MPS peak and 95% MPSR peak are plotted as the histograms in **Fig.7**. Among different head models, the 95th percentile of history needed varied from 23 ms to 37 ms for 95% MPS peak and from 20 ms to 31 ms for 95% MPSR, and longer history was needed for larger brains (**Fig.7**, **Table.2**). The time range needed for accurate calculation is given in **Fig.8**. The time ranges for most head models were within [-20 ms, 70 ms]. In a few impacts with late peak time, because of the accumulation of numerical error in long time simulation with the original data, the relative error of 95% MPS peak was larger than $\epsilon_{\text{crit}}$=6.8% when 60 ms history was included. Therefore, these cases were not plotted in **Fig.7A. Fig.8A** gives the peak time for these cases.

## Discussion

In this study, we investigated how the time window of instrumented mouthguard measurements influences the further analysis of 95% MPS and 95% MPSR in 118 American football head impacts. Different from laboratory testing, where there is only one spike near the device triggering [30], on-field data always has multiple spikes (**Fig.4**). Most peaks of kinematics (**Fig.3**) and brain strain (**Fig.5**) occur soon after the device triggering, and a few impacts had peaks much earlier or later than the triggering. It was also found that peak time for 95% MPS and 95% MPSR are influenced by brain geometry (**Fig.5**). To address this, seven head models, including the original KTH head model and six representative head models, were used in simulations to observe the effect of brain size on the kinematics. Larger brains were found to require a longer history of kinematics for accurate calculation (**Fig.7**). This is likely due to the increase in inertia and the increased distances stress waves are required to travel in larger brains during similar impact conditions. When considering all head models, a time window of [-20 ms, 70 ms] was found to give 95% MPS and 95% MPSR peaks that were non-significantly different from the original data (**Fig.6**). Therefore, [-20 ms, 70 ms] is suggested as the minimum time window for instrumented mouthguards. Additionally, head kinematics outside [-20 ms, 70 ms] were needed in a few analyses (**Fig.8**), so wider time windows will help to comprehensively collect data.

The instrumented mouthguard in the present study is triggered by linear acceleration collected by an accelerometer, and subsequently records head kinematics in a specified time window as an impact event. It should be noted that linear acceleration varies with location for a rigid body. In the MiG2.0 mouthguard, the accelerometer is located at the incisor (**Fig.1A**) in consideration of reducing the influence of jaw slamming [27]. Therefore, in this study, t=0 ms represents when the linear acceleration at the incisor reaches the triggering condition. In other mouthguards with an accelerometer at the molar [13, 30] and other types of devices [5, 9, 22, 37, 41-43, 45, 53], the angular velocity and angular acceleration may shift because t=0 ms corresponds to different triggering conditions. Therefore, the adequate pre-trigger time and post-trigger time may shift in other devices, but the length of the time window should be the same.

Accelerometers are used to trigger instrumented mouthguard sensors instead of gyroscopes as they measure angular velocity, and high angular velocity of the head may occur without an impact [17]. As a result, high-level angular acceleration may occur much earlier or later than the triggering. Since head rotation deforms the brain [19, 20], 95% MPS peak and 95% MPSR peak may also occur much earlier or later than the triggering. Early peak times can be found in kinematic parameters

(**Fig.3**), 95% MPS, and 95% MPSR (**Fig.5**). Peaks of linear acceleration at CoG were also found before the triggering (**Fig.3C**) because of the transformation from accelerometer (at the incisor) to the head CoG. Furthermore, because brain tissue is history-dependent [11, 23], head kinematics before the peak time are needed to calculate an accurate peak value. Therefore, the pre-trigger time should be long enough for the analysis of 95% MPS and 95% MPSR. Late peak time can also be found in kinematic parameters (**Fig. 3**), 95% MPS, and 95% MPSR (**Fig.5**). An example with late peak time is shown in **Fig.4A4**, where two spikes of angular acceleration far from each other are found. The linear acceleration associated with the first angular acceleration spike triggered the mouthguard and the second spike is the highest, which made 95% MPS and 95% MPSR reach peaks at the second spike in most head models (**Fig.4B4, 4C4**). Multiple instances of head contact within a short time period may be a potential explanation for this. For example, it is possible that two head impacts caused two peaks in **Fig.5A4**, respectively, and the kinematics resulting from each contact overlapped. However, the frame rate, resolution and angle of video recordings does not allow us to closely investigate this, and this may be answered by future laboratory reconstruction studies. As a result, the entire variation of kinematics should be included to calculate 95% MPS and 95% MPSR. A potential alternative of long post-trigger time is to have two subsequent recorded events and merge the head kinematics for the analysis. However, in the current device design, the sensors and microcontroller need several milliseconds to save data and reset for the next event, and the kinematics during this time will be missed. As shown in **Fig.8**, it is possible that the missing kinematics are needed to accurately calculate brain strain, so this approach may either miss the peak or cause inaccuracy.

As shown in **Fig.6**, the time window of [-20 ms, 70 ms] was found to give non-significantly different brain strain when compared with the original time window of [-50 ms, 150 ms]. However, the head kinematics outside [-20 ms, 70 ms] were needed to calculate the accurate value for brain strain in a few impacts (**Fig.8**). These impacts were found to only have relatively low brain strain and strain rate (**Fig.5**), but this may be owing to the small number of impacts. As we expect more data will be collected by MiG2.0 mouthguards in the future, the number of peaks that happen outside [-20 ms, 70 ms] should be examined.

In the present study, we observed multiple spikes within a single head impact, and found that the kinematics peak, 95% MPS peak, and 95% MPSR peak may correspond to different spikes. For example, as shown in **Fig.4A4, 4B4, 4C4**, the angular acceleration reached its peak at the second spike, but the 95% MPS for the largest male head model reached its peak at the first spike. This phenomenon shows that the whole variation of angular velocity should be recorded and raises concerns about some brain injury metrics [23, 46, 50, 51, 58] and machine-learning head models [60] that predict brain strain based on the peak values of kinematic parameters. When these metrics are applied to head impacts with multiple spikes, the kinematics at one spike may be used to predict brain strain at another, thus decreasing the prediction accuracy. Furthermore, multiple spikes will also increase the duration that brain strain sustains at a high level. Comparing impacts 1 and 2 in **Fig.4**, although the 95% MPS peaks were close, three spikes can be found in impact 2 while only one can be found in impact 1. As a result, 95% MPS in impact 2 sustains at a high level longer than in impact 1. Although we use the peak values of brain strain and strain rate to evaluate the risk of injury, we do not know if injury accumulates with long duration. Recently, repetitive head impacts have been suggested to cause more serious problems [1, 10, 32, 34, 36], which indicates that injury accumulates with multiple impacts. Therefore, the effect of multiple spikes and the duration of high brain strain and strain rate should be further investigated.

Brain geometry has been found to influence brain displacement [61], and thus brain strain as well. Recently, with the development of subject-specific head models [29], it is clear that brain geometry will influence both the peak value and the peak time of brain strain. As previously mentioned, multiple spikes of kinematics exist in one on-field head impact, and the one causing the largest 95% MPS and 95% MPSR varies among brain geometry. The history needed for accurate calculation also varies among brain geometry. Based on **Table.2 and Fig.7**, larger head models need longer histories of kinematics for calculation of both 95% MPS peak and 95% MPSR peak. This also explains why the differences of 95% MPSR peak were more significant in the largest male, 50th percentile male, and original head models (**Fig.6B1, B2, B3**) than others (**Fig.6B4, B5, B6, B7**) when the pre-trigger time was shortened to -10 ms. Therefore, the peak time varies among different head models as shown in **Figs.4, 5**. Considering that the time windows of mouthguard measurements should be suitable for all athletes, we chose the time window of [-20 ms, 70 ms], which will still avoid causing significant differences from the original time window of data for all head models. Furthermore, the different responses of 95% MPS and 95% MPSR also indicate that a subject-specific brain injury metric is warranted. The 95% MPS peak given by FE head model was used to decide the parameters of brain injury metrics [12, 14, 28, 49]. However, as shown in **Figs.3, 4**, the 95% MPS peak in the largest male head model is much larger than that in the smallest female. As a result, the current injury threshold may miss some actual injury cases.

This study has several limitations. First, although different head models were adopted, the individual differences in the brain tissue properties were not considered. The material model and parameters will influence the history needed for accurate calculation, and then influence the time window. In the KTH model [23], the Ogden hyperelasticity model [39] was assumed to model the brain tissue, and the parameters were obtained from experiments [11]. With more knowledge of the individual differences in brain tissue, future research should be performed to include the influence of material properties on necessary time windows. Regarding brain geometry, scaling the model according to moment of inertia may not accurately represent the influence of individual differences. We expect the development of method of mesh morphology or remeshing to adopt the individual difference in the future. Second, the history needed for an accurate 95% MPS peak was not calculated in a few cases (**Fig.8A**) because the relative error was higher than 6.8% although 60 ms history was included. This might be caused by the accumulation of numerical error after long-time simulations and need further investigation. Lastly, the 118 head impacts used in this study were collected by the MiG2.0 mouthguard, which has a time window of [-50 ms, 150 ms]. To the best of our knowledge, this is the longest time window of any head impact sensor, and we found that data shortened to a time window of [-20 ms, 70 ms] will give results that are not significantly different from the original time window. However, it is possible that this dataset does not reflect the actual distribution of peak times in American football, as peaks may still occur before or after the [-50 ms, 150 ms] time window.

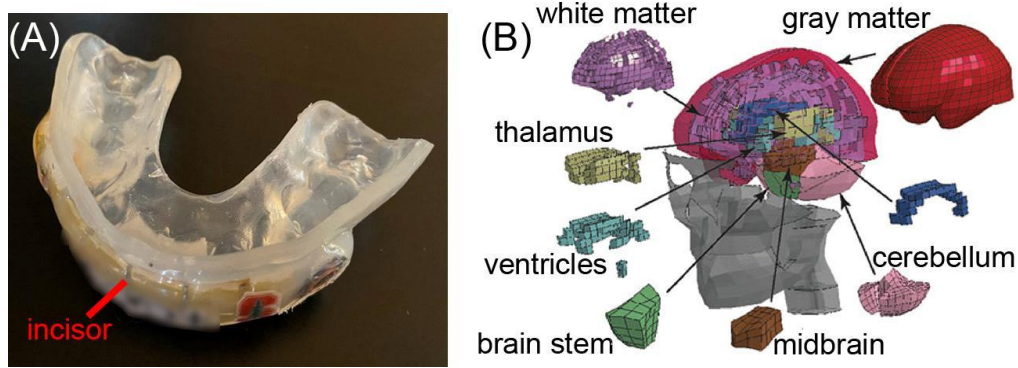

Figure 1. (A) Stanford Instrumented Mouthguard (MiG2.0) [31]. The location of the incisor is shown. (B) KTH finite element head model [23].

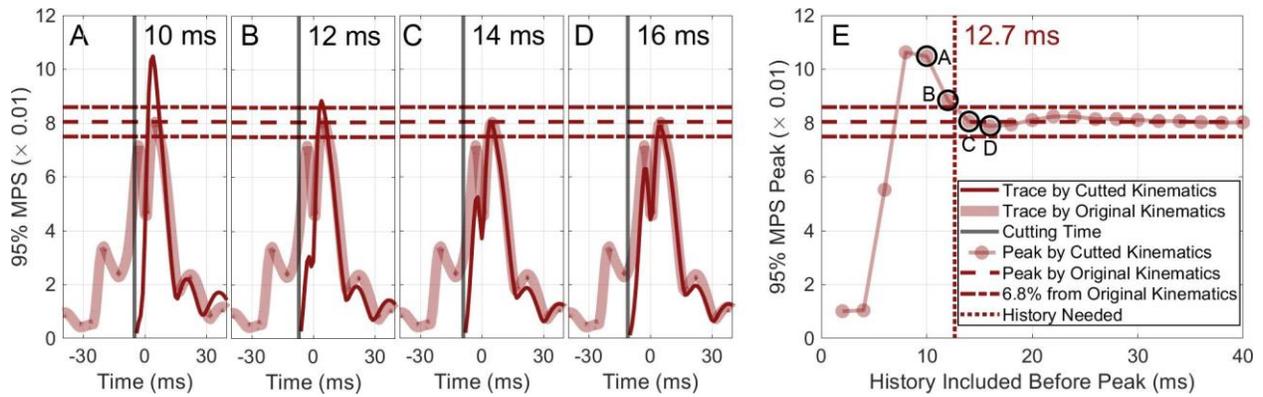

Figure 2. Example calculation of the history needed for 95% MPS peak. (A-D) Trace of 95% MPS calculated by the kinematics shortened at 10 ms, 12 ms, 14 ms, 16 ms before the peak time, respectively. (E) 95% MPS peak calculated by the kinematics shortened at every 2 ms to 40 ms before the peak time. (A-D) is indicated in (E), and the history needed is calculated by cases B and C.

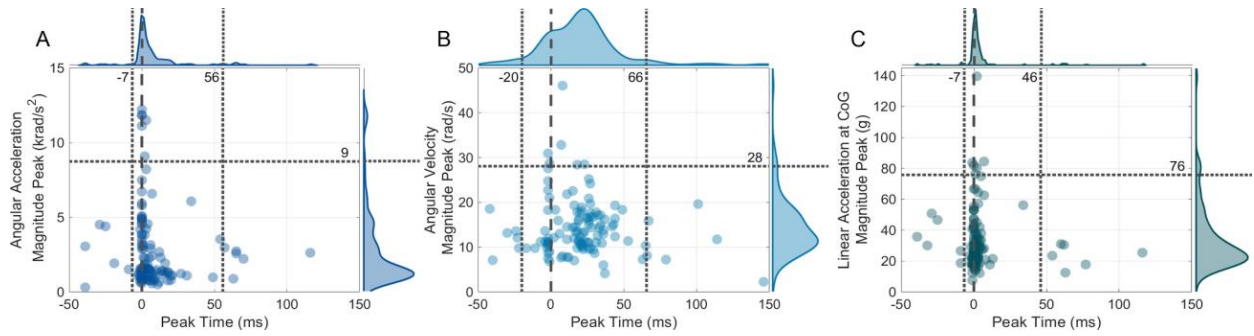

Figure 3. The head kinematics peaks and the time for the peaks of 118 on-field football impacts. The vertical dashed line is the trigger time (t=0 ms). The left and right vertical dotted lines are the 5th percentile and the 95th percentile of the time for the peaks, respectively, and the horizontal dotted line is the 95th percentile of the head kinematics peaks.

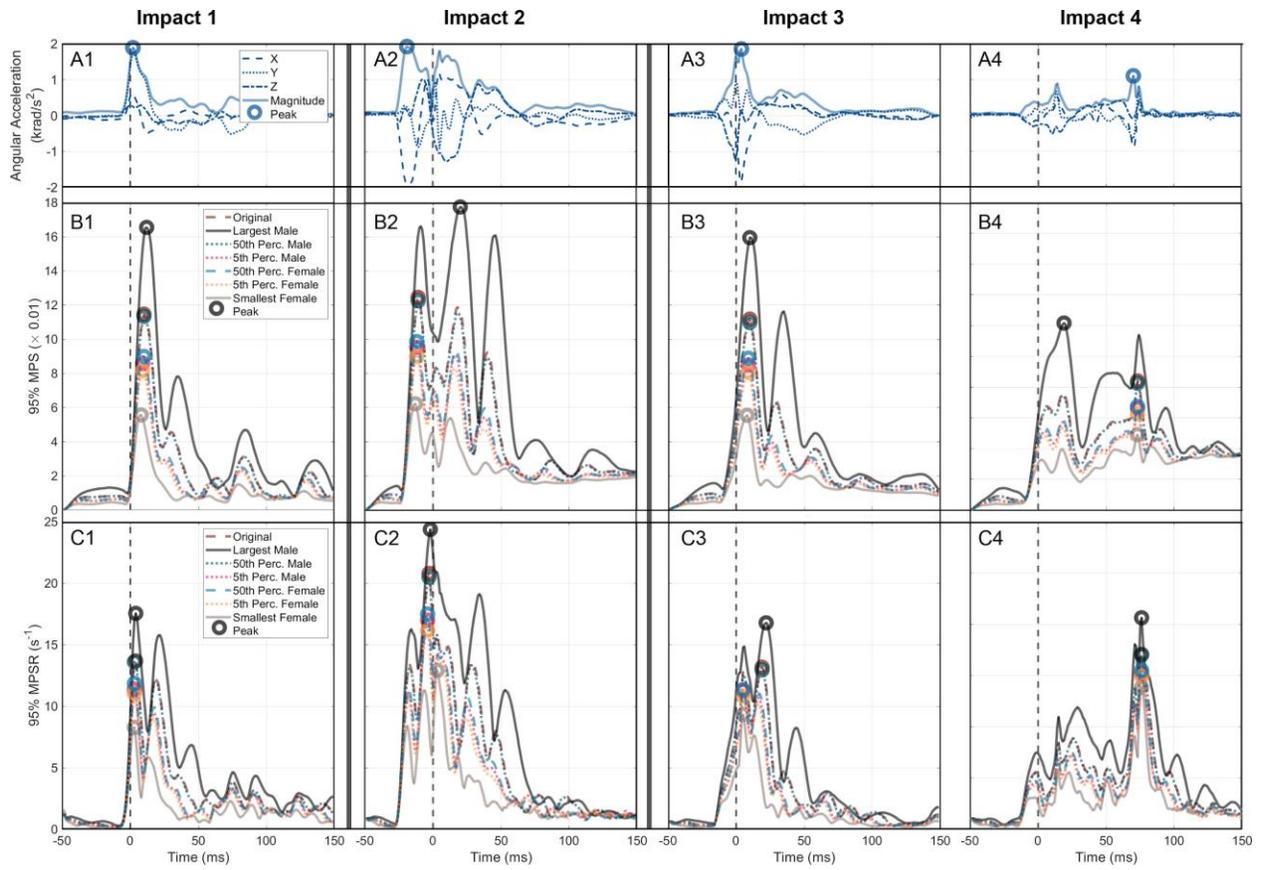

Figure 4. Four example head impacts. (A) angular acceleration components and magnitude (X: back to front, Y: left to right, Z: top to bottom); (B) 95% MPS calculated by seven heads models; (C) 95% MPSR calculated by seven heads models.

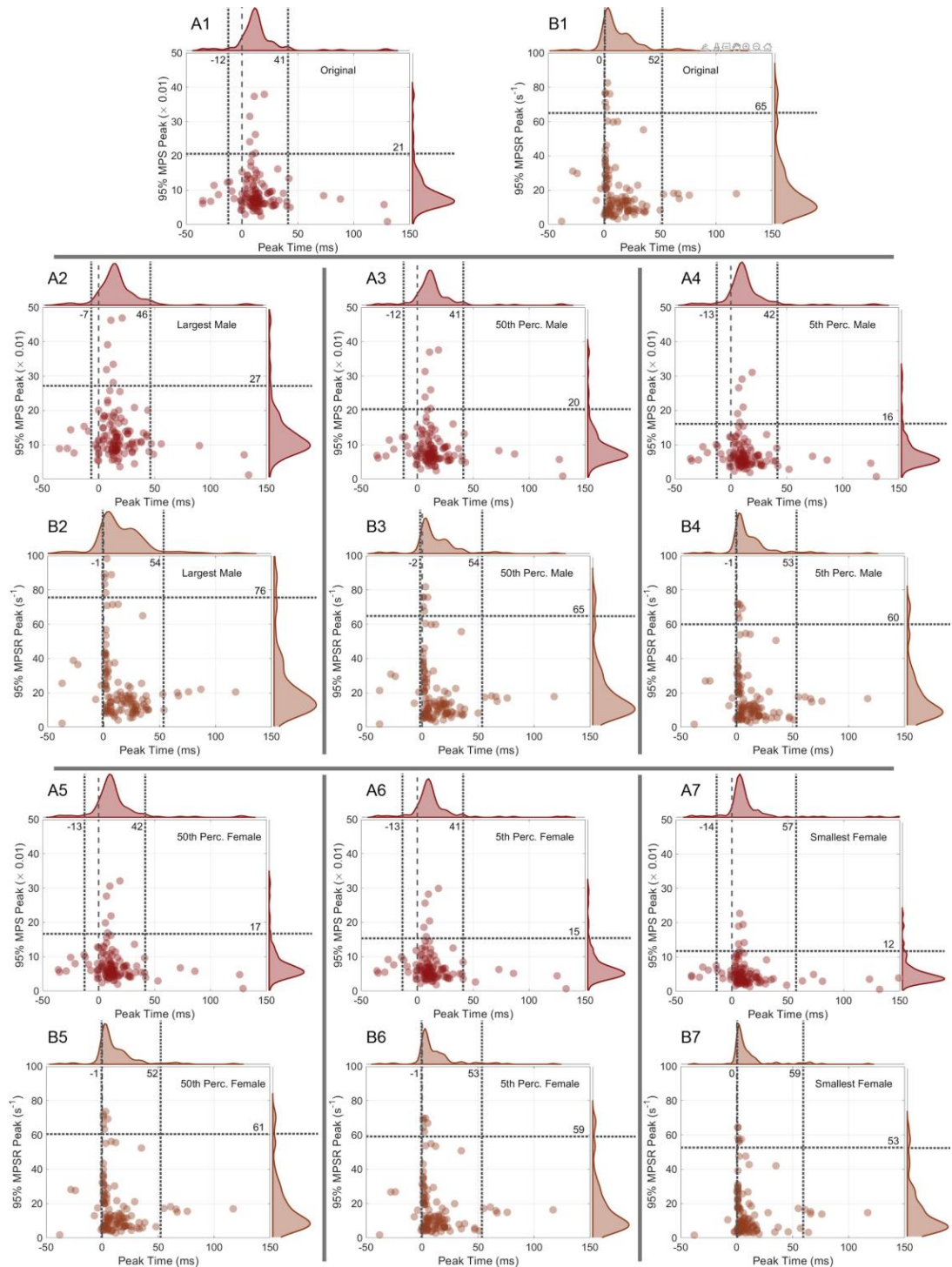

Figure 5. (A) The 95% MPS peaks and the time for the peaks; (B) the 95% MPSR peaks and the time for the peaks. (1-7) KTH original and six representative brains. The vertical dashed line was the trigger time (t=0 ms). The left and right vertical dotted lines were the 5th percentile and the 95th percentile of the time for the peaks, respectively, and the horizontal dotted line is the 95th percentile of the 95% MPS or MPSR peaks.

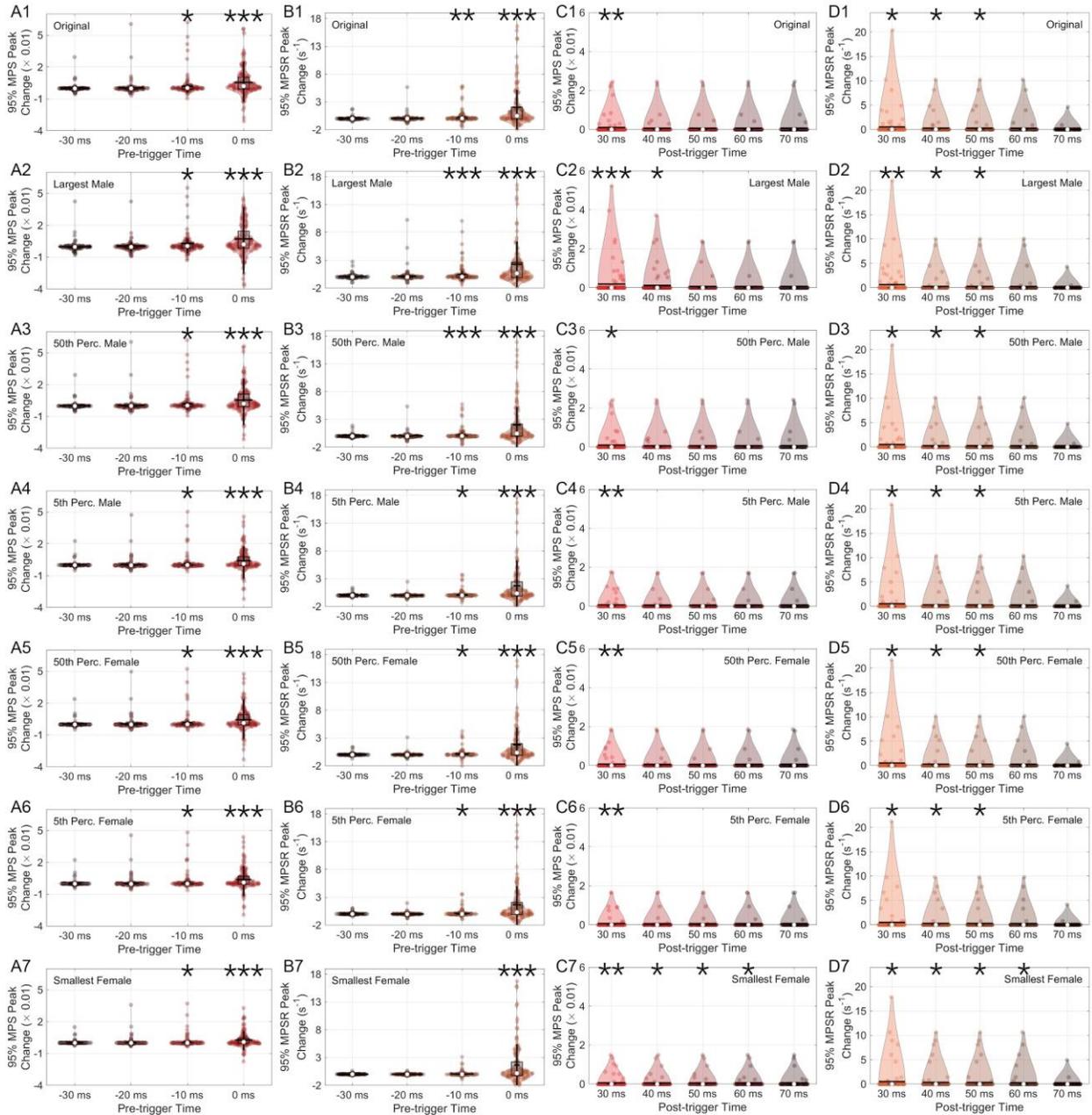

Figure 6. Changes of 95% MPS and 95% MPSR peaks caused by the pre-trigger and post-trigger time (the results by original data minus the results by new data with different time windows). (A) 95% MPS peak and (B) 95% MPSR peak calculated by different pre-trigger times; (C) 95% MPS peak and (D) 95% MPSR peak calculated by different post-trigger times. (1-7) KTH original and six representative brains. Each group (violin) is compared with the original data (A, B: original pre-trigger time: -50 ms; C, D: original post-trigger time: 150 ms) by pairwise t-test. (*:$p<0.05$, **:$p<0.01$, ***:$p<0.005$).

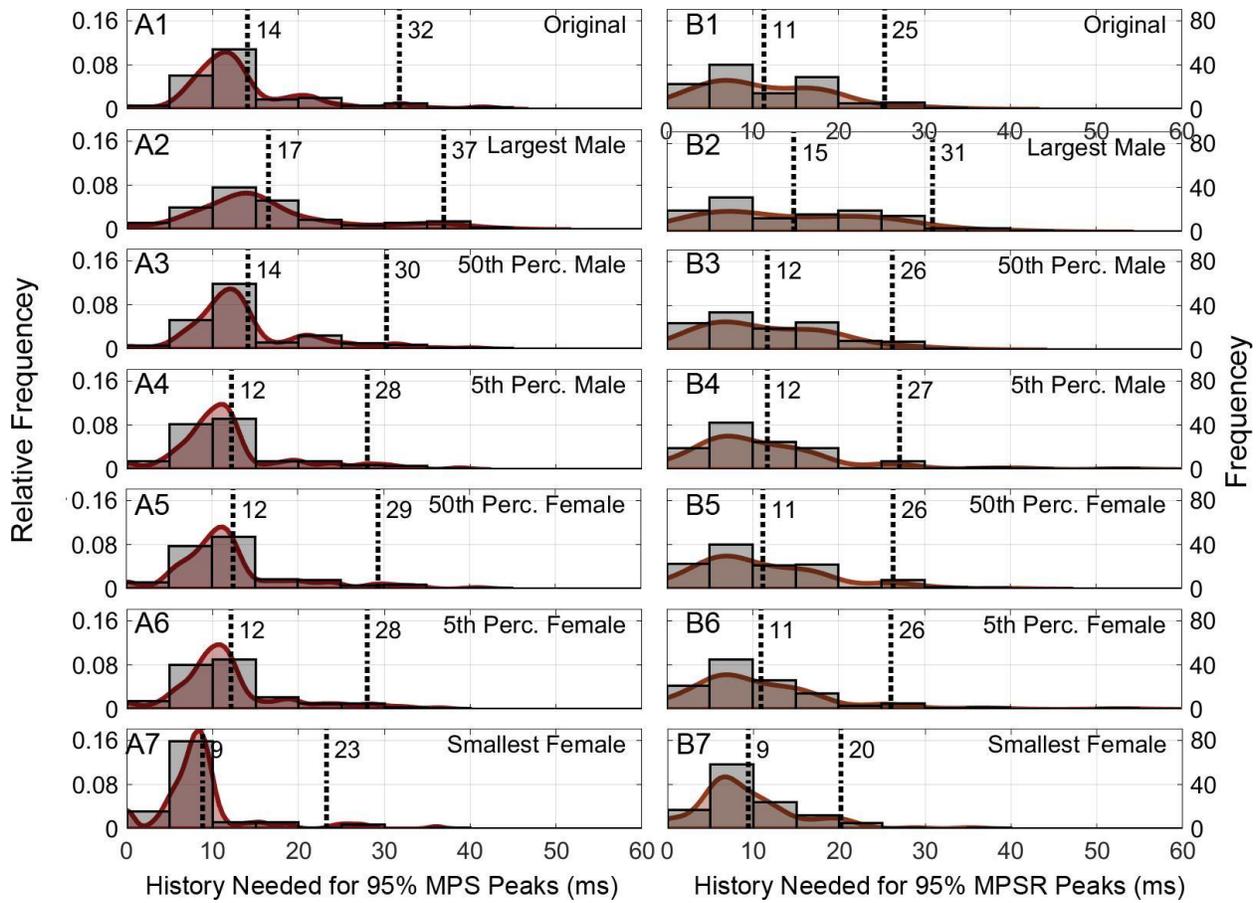

Figure 7. Histogram of the history needed for accurate calculation (within critical relative error of 6.8%) of 95% MPS peak (A) and 95% MPSR peak (B). (1-7) KTH original and six representative brains. In each plot, the curves are the kernel density, and the left and right vertical dotted lines are the mean and the 95th percentile history needed.

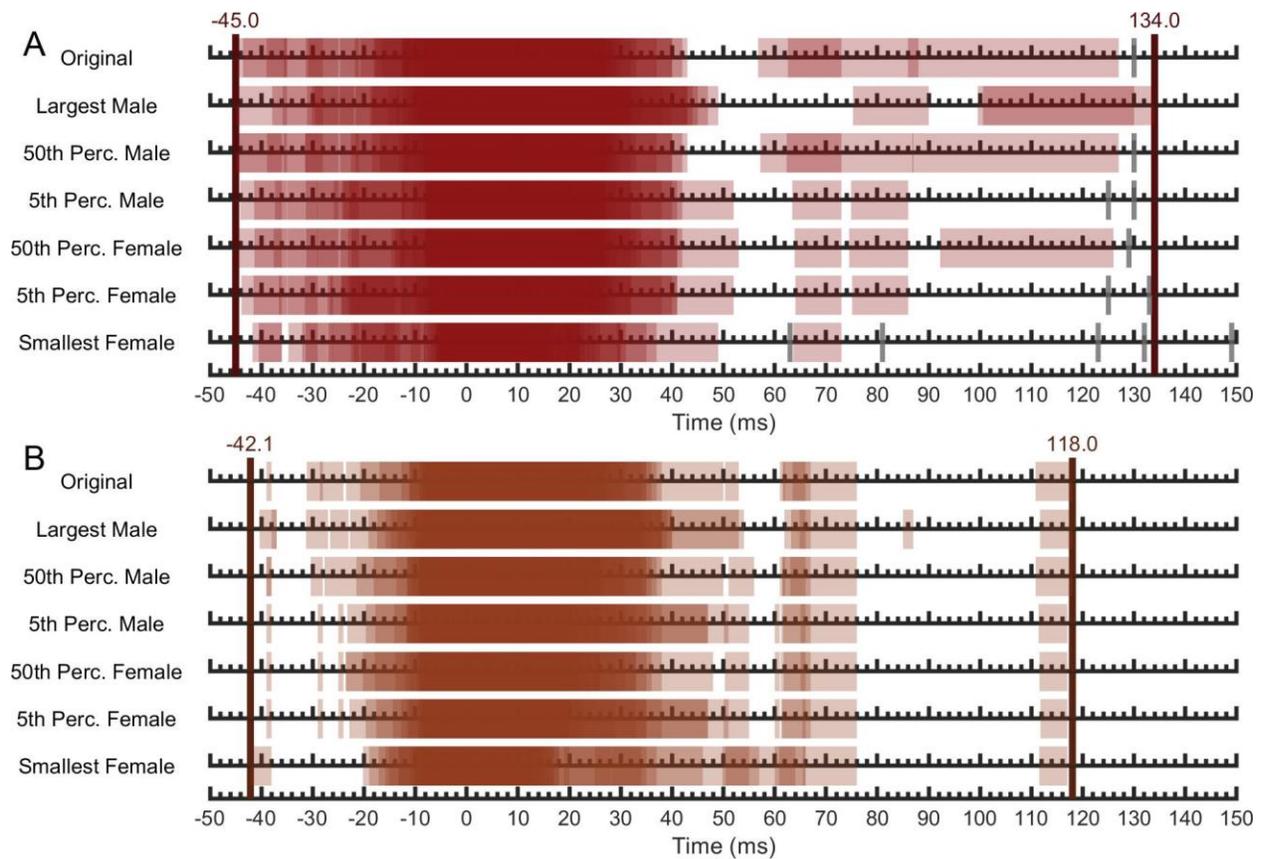

Figure 8. Time ranges for the accurate calculation (within $\epsilon_i$ of 6.8%) of 95% MPS peak (A) and 95% MPSR peak (B) in 118 on-field football impacts. KTH original and six representative brains shown at the left. The left and right vertical lines are the earliest and the latest time needed, respectively. The gray bars in (A) are the cases in which the relative error of 95% MPS peak was larger than $\epsilon_i$ even when 60 ms history was included.

Table 1. Time window of wearable devices to measure head kinematics. t=0 ms is the device triggering.

| Devices | Time Windows (ms) |
|---|---|
| Head Impact Telemetry System (HITS) | [-8, 32] [45, 53]<br>[-12, 28] [9] |
| GForce Tracker (sensor module for helmet) | [-8, 32] [21] |
| SIM-G (sensor module for headband) | [-10, 52] [21] |
| X-patch | [-10, 90] [43] |
| Instrumented Mouthguard | [-10, 50] [13] |
| Instrumented Mouthguard | [-15, 45] [44] |
| Vector Mouthguard | [-16, 80] [21] |
| Sports & Wellbeing Analysis Mouthguard | [-1, 101] [30] |
| Prevent Mouthguard | [-10, 40] [21, 30] |
| Stanford Instrument Mouthguard (MiG) | [-50, 150] [30] |

Table 2. Scaling coefficient from original KTH head model to six representative head models based on moment of inertia at each axis [54]. The frame of reference is: X, back to front, Y: left to right; Z, top to bottom. (Perc.: Percentile)

| Head Model | Subject ID | X | Y | Z |
|---|---|---|---|---|
| Largest Male | 223929 | 1.1374 | 1.2999 | 1.1967 |
| 50th Perc. Male | 172635 | 0.9770 | 1.0122 | 0.9765 |
| 5th Perc. Male | 185038 | 0.8850 | 0.8900 | 0.8511 |
| 50th Perc. Female | 771354 | 0.8673 | 0.9068 | 0.8872 |
| 5th Perc. Female | 568963 | 0.8176 | 0.8565 | 0.8526 |
| Smallest Female | 118124 | 0.7035 | 0.7287 | 0.6959 |

# References

1. Bailes J. E., A. L. Petraglia, B. I. Omalu, E. Nauman and T. Talavage. Role of subconcussion in repetitive mild traumatic brain injury: a review. *Journal of neurosurgery* 119: 1235-1245, 2013.
2. Bartsch A., S. Samorezov, E. Benzel, V. Miele and D. Brett. Validation of an "intelligent mouthguard" single event head impact dosimeter. SAE Technical Paper, 2014.
3. Bartsch A. J., D. Hedin, J. Alberts, E. C. Benzel, J. Cruickshank, R. S. Gray, K. Cameron, M. N. Houston, T. Rooks and G. McGinty. High energy side and rear American football head impacts cause obvious performance decrement on video. *Annals of Biomedical Engineering* 1-11, 2020.
4. Bridgman H., M. T. Kwong and J. H. Bergmann. Mechanical safety of embedded electronics for in-body wearables: A smart mouthguard study. *Annals of Biomedical Engineering* 47: 1725-1737, 2019.
5. Caccese J. B., L. C. Lamond, T. A. Buckley and T. W. Kaminski. Reducing purposeful headers from goal kicks and punts may reduce cumulative exposure to head acceleration. *Research in Sports Medicine* 24: 407-415, 2016.
6. Camarillo D. B., P. B. Shull, J. Mattson, R. Shultz and D. Garza. An instrumented mouthguard for measuring linear and angular head impact kinematics in American football. *Annals of Biomedical Engineering* 41: 1939-1949, 2013.
7. Dewan M. C., A. Rattani, S. Gupta, R. E. Baticulon, Y.-C. Hung, M. Punchak, A. Agrawal, A. O. Adeleye, M. G. Shrime and A. M. Rubiano. Estimating the global incidence of traumatic brain injury. *Journal of neurosurgery* 130: 1080-1097, 2018.
8. Domel A. G., S. J. Raymond, C. Giordano, Y. Liu, S. A. Yousefsani, M. Fanton, I. Pirozzi, A. Kight, B. Avery and A. Boumis. A New Open-Access Platform for Measuring and Sharing mTBI Data. *arXiv preprint arXiv:2010.08485* 2020.
9. Duma S. M., S. J. Manoogian, W. R. Bussone, P. G. Brolinson, M. W. Goforth, J. J. Donnenwerth, R. M. Greenwald, J. J. Chu and J. J. Crisco. Analysis of real-time head accelerations in collegiate football players. *Clinical Journal of Sport Medicine* 15: 3-8, 2005.
10. Fehily B. and M. Fitzgerald. Repeated mild traumatic brain injury: potential mechanisms of damage. *Cell transplantation* 26: 1131-1155, 2017.
11. Franceschini G., D. Bigoni, P. Regitnig and G. A. Holzapfel. Brain tissue deforms similarly to filled elastomers and follows consolidation theory. *Journal of the Mechanics and Physics of Solids* 54: 2592-2620, 2006.
12. Gabler L. F., J. R. Crandall and M. B. Panzer. Development of a second-order system for rapid estimation of maximum brain strain. *Annals of Biomedical Engineering* 47: 1971-1981, 2019.
13. Gabler L. F., S. H. Huddleston, N. Z. Dau, D. J. Lessley, K. B. Arbogast, X. Thompson, J. E. Resch and J. R. Crandall. On-Field Performance of an Instrumented Mouthguard for Detecting Head Impacts in American Football. *Annals of Biomedical Engineering* 1-14, 2020.
14. Gabler L. F., H. Joodaki, J. R. Crandall and M. B. Panzer. Development of a single-degree-of-freedom mechanical model for predicting strain-based brain injury responses. *Journal of biomechanical engineering* 140: 2018.
15. Hajiaghamemar M. and S. S. Margulies. Multi-scale white matter tract embedded brain finite element model predicts the location of traumatic diffuse axonal injury. *Journal of Neurotrauma* 2020.
16. Hardy W. N., C. D. Foster, M. J. Mason, K. H. Yang, A. I. King and S. Tashman. Investigation of head injury mechanisms using neutral density technology and high-speed biplanar X-ray. SAE Technical Paper, 2001.
17. Hernandez F. and D. B. Camarillo. Voluntary head rotational velocity and implications for brain injury risk metrics. *Journal of Neurotrauma* 36: 1125-1135, 2019.

18. Hernandez F., L. C. Wu, M. C. Yip, K. Laksari, A. R. Hoffman, J. R. Lopez, G. A. Grant, S. Kleiven and D. B. Camarillo. Six degree-of-freedom measurements of human mild traumatic brain injury. *Annals of Biomedical Engineering* 43: 1918-1934, 2015.
19. Holbourn A. Mechanics of head injuries. *The Lancet* 242: 438-441, 1943.
20. Ji S., W. Zhao, Z. Li and T. W. McAllister. Head impact accelerations for brain strain-related responses in contact sports: a model-based investigation. *Biomechanics and Modeling in Mechanobiology* 13: 1121-1136, 2014.
21. Kieffer E. E., M. T. Begonia, A. M. Tyson and S. Rowson. A two-phased approach to quantifying head impact sensor accuracy: in-laboratory and on-field assessments. *Annals of Biomedical Engineering* 48: 2613-2625, 2020.
22. King D., P. Hume, C. Gissane and T. Clark. Head impacts in a junior rugby league team measured with a wireless head impact sensor: an exploratory analysis. *Journal of Neurosurgery: Pediatrics* 19: 13-23, 2017.
23. Kleiven S. Predictors for traumatic brain injuries evaluated through accident reconstructions. SAE Technical Paper, 2007.
24. Knox T. Validation of earplug accelerometers as a means of measuring head motion. SAE Technical Paper, 2004.
25. Kuo C., L. Wu, J. Loza, D. Senif, S. C. Anderson and D. B. Camarillo. Comparison of video-based and sensor-based head impact exposure. *PloS one* 13: e0199238, 2018.
26. Kuo C., L. Wu, W. Zhao, M. Fanton, S. Ji and D. B. Camarillo. Propagation of errors from skull kinematic measurements to finite element tissue responses. *Biomechanics and Modeling in Mechanobiology* 17: 235-247, 2018.
27. Kuo C., L. C. Wu, B. T. Hammoor, J. F. Luck, H. C. Cutcliffe, R. C. Lynall, J. R. Kait, K. R. Campbell, J. P. Mihalik, C. R. Bass and D. B. Camarillo. Effect of the mandible on mouthguard measurements of head kinematics. *Journal of Biomechanics* 49: 1845-1853, 2016.
28. Laksari K., M. Fanton, L. C. Wu, T. H. Nguyen, M. Kurt, C. Giordano, E. Kelly, E. O'Keeffe, E. Wallace and C. Doherty. Multi-directional dynamic model for traumatic brain injury detection. *Journal of Neurotrauma* 37: 982-993, 2020.
29. Li X., Z. Zhou and S. Kleiven. An anatomically detailed and personalizable head injury model: Significance of brain and white matter tract morphological variability on strain. *Biomechanics and Modeling in Mechanobiology* 1-29, 2020.
30. Liu Y., A. G. Domel, S. A. Yousefsani, J. Kondic, G. Grant, M. Zeineh and D. B. Camarillo. Validation and comparison of instrumented mouthguards for measuring head kinematics and assessing brain deformation in football impacts. *Annals of Biomedical Engineering* 48: 2580-2598, 2020.
31. Liu Y., X. Zhan, A. G. Domel, M. Fanton, Z. Zhou, S. J. Raymond, H. V. Alizadeh, N. J. Cecchi, M. Zeineh and G. Grant. Theoretical and numerical analysis for angular acceleration being determinant of brain strain in mTBI. *arXiv preprint arXiv:2012.13507* 2020.
32. Meabon J. S., B. R. Huber, D. J. Cross, T. L. Richards, S. Minoshima, K. F. Pagulayan, G. Li, K. D. Meeker, B. C. Kraemer and E. C. Petrie. Repetitive blast exposure in mice and combat veterans causes persistent cerebellar dysfunction. *Science translational medicine* 8: 321ra326-321ra326, 2016.
33. Mihalik J. P., A. Chandran, J. R. Powell, P. R. Roby, K. M. Guskiewicz, B. D. Stemper, A. S. Shah, S. Rowson, S. Duma and J. Harezlak. Do head injury biomechanics predict concussion clinical recovery in college American football players? *Annals of Biomedical Engineering* 48: 2555-2565, 2020.
34. Miller L. E., J. E. Urban, E. M. Davenport, A. K. Powers, C. T. Whitlow, J. A. Maldjian and J. D. Stitzel. Brain Strain: Computational Model-Based Metrics for Head Impact Exposure and Injury Correlation. *Annals of Biomedical Engineering* 1-14, 2020.
35. Motiwale S., W. Eppler, D. Hollingsworth, C. Hollingsworth, J. Morgenthau and R. H. Kraft. Application of neural networks for filtering non-impact transients recorded from biomechanical sensors. In: *2016 IEEE-EMBS International Conference on Biomedical and Health Informatics (BHI)*IEEE, 2016, p. 204-207.
36. Mouzon B. C., C. Bachmeier, J. O. Ojo, C. M. Acker, S. Ferguson, D. Paris, G. Ait-Ghezala, G. Crynen, P. Davies and M. Mullan. Lifelong behavioral and neuropathological consequences of repetitive mild traumatic brain injury. *Annals of clinical and translational neurology* 5: 64-80, 2018.
37. O'Day K. M., E. M. Koehling, L. R. Vollavanh, D. Bradney, J. M. May, K. M. Breedlove, E. L. Breedlove, P. Blair, E. A. Nauman and T. G. Bowman. Comparison of head impact location during games and practices in Division III men's lacrosse players. *Clinical Biomechanics* 43: 23-27, 2017.
38. O'Keeffe E., E. Kelly, Y. Liu, C. Giordano, E. Wallace, M. Hynes, S. Tiernan, A. Meagher, C. Greene and S. Hughes. Dynamic Blood–Brain Barrier Regulation in Mild Traumatic Brain Injury. *Journal of Neurotrauma* 37: 347-356, 2020.
39. Ogden R. W. *Non-linear elastic deformations*. Courier Corporation, 1997.
40. Ozga J. E., J. M. Povroznik, E. B. Engler-Chiurazzi and C. V. Haar. Executive (dys) function after traumatic brain injury: special considerations for behavioral pharmacology. *Behavioural pharmacology* 2018.
41. Panzer M. B., R. Cameron, R. S. Salzar, J. Pellettiere and B. Myers. Evaluation of ear-mounted sensors for determining impact head acceleration. *Shock* 1: 54.54, 2009.
42. Pellman E. J., D. C. Viano, A. M. Tucker and I. R. Casson. Concussion in professional football: Location and direction of helmet impacts—Part 2. *Neurosurgery* 53: 1328-1341, 2003.
43. Press J. N. and S. Rowson. Quantifying head impact exposure in collegiate women's soccer. *Clinical Journal of Sport Medicine* 27: 104-110, 2017.
44. Rich A. M., T. M. Filben, L. E. Miller, B. T. Tomblin, A. R. Van Gorkom, M. A. Hurst, R. T. Barnard, D. S. Kohn, J. E. Urban and J. D. Stitzel. Development, validation and pilot field deployment of a custom mouthpiece for head impact measurement. *Annals of Biomedical Engineering* 47: 2109-2121, 2019.
45. Rowson S., J. G. Beckwith, J. J. Chu, D. S. Leonard, R. M. Greenwald and S. M. Duma. A six degree of freedom head acceleration measurement device for use in football. *Journal of applied biomechanics* 27: 8-14, 2011.
46. Rowson S. and S. M. Duma. Brain injury prediction: assessing the combined probability of concussion using linear and rotational head acceleration. *Annals of Biomedical Engineering* 41: 873-882, 2013.
47. Siegmund G. P., K. M. Guskiewicz, S. W. Marshall, A. L. DeMarco and S. J. Bonin. A headform for testing helmet and mouthguard sensors that measure head impact severity in football players. *Annals of Biomedical Engineering* 42: 1834-1845, 2014.
48. Siegmund G. P., K. M. Guskiewicz, S. W. Marshall, A. L. DeMarco and S. J. Bonin. Laboratory validation of two wearable sensor systems for measuring head impact severity in football players. *Annals of Biomedical Engineering* 44: 1257-1274, 2016.
49. Takahashi Y. and T. Yanaoka. A study of injury criteria for brain injuries in traffic accidents. In: *25th International Technical Conference on the Enhanced Safety of Vehicles (ESV) National Highway Traffic Safety Administration*2017.
50. Takhounts E. G., M. J. Craig, K. Moorhouse, J. McFadden and V. Hasija. Development of brain injury criteria (BrIC). SAE Technical Paper, 2013.
51. Takhounts E. G., V. Hasija, S. A. Ridella, S. Rowson and S. M. Duma. Kinematic rotational brain injury criterion (BRIC). In: *Proceedings of the 22nd enhanced safety of vehicles conference. Paper*2011, p. 1-10.
52. Tiernan S., A. Meagher, D. O'Sullivan, E. O'Keeffe, E. Kelly, E. Wallace, C. P. Doherty, M. Campbell, Y. Liu and A. G. Domel. Concussion and the severity of head impacts in mixed martial arts. *Proceedings of the Institution of Mechanical Engineers, Part H: Journal of Engineering in Medicine* 0954411920947850, 2020.

53. Urban J. E., E. M. Davenport, A. J. Golman, J. A. Maldjian, C. T. Whitlow, A. K. Powers and J. D. Stitzel. Head impact exposure in youth football: high school ages 14 to 18 years and cumulative impact analysis. *Annals of Biomedical Engineering* 41: 2474-2487, 2013.
54. Van Essen D. C., S. M. Smith, D. M. Barch, T. E. Behrens, E. Yacoub, K. Ugurbil and W.-M. H. Consortium. The WU-Minn human connectome project: an overview. *Neuroimage* 80: 62-79, 2013.
55. Wu L. C., C. Kuo, J. Loza, M. Kurt, K. Laksari, L. Z. Yanez, D. Senif, S. C. Anderson, L. E. Miller and J. E. Urban. Detection of American football head impacts using biomechanical features and support vector machine classification. *Scientific Reports* 8: 1-14, 2017.
56. Wu L. C., K. Laksari, C. Kuo, J. F. Luck, S. Kleiven, R. Cameron and D. B. Camarillo. Bandwidth and sample rate requirements for wearable head impact sensors. *Journal of Biomechanics* 49: 2918-2924, 2016.
57. Wu L. C., V. Nangia, K. Bui, B. Hammoor, M. Kurt, F. Hernandez, C. Kuo and D. B. Camarillo. In vivo evaluation of wearable head impact sensors. *Annals of Biomedical Engineering* 44: 1234-1245, 2016.
58. Yanaoka T., Y. Dokko and Y. Takahashi. Investigation on an injury criterion related to traumatic brain injury primarily induced by head rotation. SAE Technical Paper, 2015.
59. Zhan X., Y. Li, Y. Liu, A. G. Domel, H. V. Alidazeh, S. J. Raymond, J. Ruan, S. Barbat, S. Tiernan and O. Gevaert. Prediction of brain strain across head impact subtypes using 18 brain injury criteria. *arXiv preprint arXiv:2012.10006* 2020.
60. Zhan X., Y. Liu, S. J. Raymond, H. V. Alizadeh, A. G. Domel, O. Gevaert, M. Zeineh, G. Grant and D. B. Camarillo. Deep Learning Head Model for Real-time Estimation of Entire Brain Deformation in Concussion. *arXiv preprint arXiv:2010.08527* 2020.
61. Zou H., S. Kleiven and J. P. Schmiedeler. The effect of brain mass and moment of inertia on relative brain–skull displacement during low-severity impacts. *International journal of crashworthiness* 12: 341-353, 2007.